\documentclass[floatfix,altaffilletter,superscriptaddress,preprintnumbers,tightenlines,showpacs,showkeys,nofootinbib]{revtex4-2}
\usepackage[utf8]{inputenc}
\usepackage[colorlinks=true,citecolor=blue,linkcolor=blue]{hyperref}
\usepackage[normalem]{ulem}
\usepackage{amsmath,amssymb, mathrsfs, physics, aas_macros}
\usepackage{epsfig}
\usepackage{graphicx}            
\usepackage{booktabs,multirow}
\usepackage{url}
\usepackage{color}
\usepackage[dvipsnames]{xcolor}
\usepackage{soul}
\usepackage[capitalise, english]{cleveref}
\usepackage{siunitx}
\usepackage{xspace}
\usepackage{enumitem}
\usepackage{fontawesome}

\newcommand\myshade{80}
\colorlet{mylinkcolor}{Black} 
\colorlet{mycitecolor}{Red}
\colorlet{myurlcolor}{violet}

\hypersetup{
  linkcolor  = mylinkcolor!\myshade!black,
  citecolor  = mycitecolor!\myshade!black,
  urlcolor   = myurlcolor!\myshade!black,
  colorlinks = true
}

\newcommand{\CL}{{\tt ${\mathcal C}$osmo${\mathcal L}$attice}}

\definecolor{cborange}{HTML}{e69f00}
\definecolor{cbgreen}{HTML}{009e73}
\definecolor{cbyellow}{HTML}{f1dd42}
\definecolor{cblblue}{HTML}{56b4e9}
\definecolor{cbblue}{HTML}{0072b2}
\definecolor{cbpurple}{HTML}{cc79a7}
\definecolor{defgrey}{HTML}{9f9f9f}
\definecolor{defgreen}{HTML}{8eba42}
\definecolor{defred}{HTML}{e24a33}
\definecolor{defblue}{HTML}{358abd}
\definecolor{defpurple}{HTML}{988ed5}

\makeatletter
\providecommand*{\diff}%
	{\@ifnextchar^{\DIfF}{\DIfF^{}}}
\def\DIfF^#1{%
	\mathop{\mathrm{\mathstrut d}}%
		\nolimits^{#1}\gobblespace}
\def\gobblespace{%
	\futurelet\diffarg\opspace}
\def\opspace{%
	\let\DiffSpace\!%
	\ifx\diffarg(%
		\let\DiffSpace\relax
	\else
		\ifx\diffarg[%
			\let\DiffSpace\relax
		\else
  			\ifx\diffarg\{%
				\let\DiffSpace\relax
			\fi\fi\fi\DiffSpace}

\newcolumntype{C}[1]{>{\centering\let\newline\\\arraybackslash\hspace{0pt}}m{#1}}

\begin{document}

\title{Ricci Reheating on the Lattice}

\author{Daniel G. Figueroa}\email{daniel.figueroa@ific.uv.es}
\affiliation{Instituto de F\'isica Corpuscular (IFIC), Consejo Superior de Investigaciones \\Cient\'ificas (CSIC) and Universitat de Val\`{e}ncia (UV), 46980 Valencia, Spain.}

\author{Toby Opferkuch}\email{toby.opferkuch@sissa.it}
\affiliation{SISSA International School for Advanced Studies, Via Bonomea 265, 34136, Trieste, Italy}
\affiliation{INFN, Sezione di Trieste, Via Bonomea 265, 34136, Trieste, Italy}

\author{Ben A. Stefanek}\email{benjamin.stefanek@kcl.ac.uk}
\affiliation{Physics Department, King’s College London, Strand, London, WC2R 2LS, United Kingdom}

\preprint{KCL-PH-TH/2024-19}

\begin{abstract}
\noindent We study the dynamics of a non-minimally coupled (NMC) scalar spectator field in non-oscillatory inflationary scenarios, where there is a transition from inflation to kination domination (KD). Engineering a realistic finite-duration transition through a CMB-compatible inflaton potential, we calculate the initial tachyonic growth of the NMC field during KD and perform lattice simulations of the subsequent non-linear dynamics. We characterize the regularization effect on the tachyonic growth, either due to self-interactions, or via gravitational backreaction when the NMC field grows to dominate the energy of the universe. Our study provides the first realistic treatment of the dynamics, with significant improvements compared to previous work, where one or more of the following aspects were assumed: ($i$) the background expansion can be neglected during the tachyonic growth, ($ii$) coherence of the NMC field, ($iii$) coherence of the inflaton, ($iv$) instantaneous transition, and ($v$) a KD equation of state of exactly $w = 1$. Using our methodology, which requires none of the above assumptions, we determine the conditions to achieve proper 
reheating, {\it i.e.~}energetic dominance of the NMC field over the inflaton. We characterize the time and energy scales of the problem, either for backreaction due to self-interactions, or (as a novelty of this work) due to gravitational effects. Finally, we calculate $\mathcal{O}(1)$ lattice correction factors to analytic scaling relations derived by some of us in previous work. This enables simple future studies without the need to run lattice simulations.
\end{abstract}

\maketitle
\tableofcontents
\section{Introduction}
\label{sec:intro}

Precious little is known about the history of our Universe predating the onset of Big-Bang nucleosynthesis (BBN). Although there is mounting evidence supporting the existence of an inflationary phase in the early Universe~\cite{Planck:2018jri}, there remains little observational information concerning the dynamics ending this phase, which must bring the universe into a state of radiation domination in order to achieve successful BBN. Therefore, one way or another, a {\it reheating} stage producing different particle species, dominated by relativistic degrees of freedom, must follow after inflation. These species must eventually thermalize and dominate the universe's energy budget, ensuring that the subsequent expansion aligns with the {\it hot Big Bang} thermal paradigm prior to BBN.

The possibilities for reheating depend largely upon the manner in which the inflaton is coupled to other particle species. At one extreme lie direct couplings allowing (non-)perturbative decays of the inflaton into quanta of various particle species~\cite{Linde:1981mu,Albrecht:1982mp,Dolgov:1982th,Abbott:1982hn,Traschen:1990sw,Kofman:1994rk,Shtanov:1994ce,Kaiser:1995fb,Kofman:1997yn,Greene:1997fu,Kaiser:1997mp,Kaiser:1997hg,Greene:1998nh,Greene:2000ew,Peloso:2000hy}, see {\it e.g.}~Refs.~\cite{Felder:2000hj,Felder:2001kt,Copeland:2002ku,GarciaBellido:2002aj,Rajantie:2000nj,Copeland:2001qw,Smit:2002yg,GarciaBellido:2003wd,Tranberg:2003gi,Skullerud:2003ki,vanderMeulen:2005sp,DiazGil:2007dy,DiazGil:2008tf,Dufaux:2010cf,Berges:2010zv,Tranberg:2017lrx,Deskins:2013lfx,Adshead:2015pva,Adshead:2016iae,Lozanov:2016hid,Lozanov:2016pac,Figueroa:2017qmv,Adshead:2017xll,Adshead:2018doq,Cuissa:2018oiw,Adshead:2019lbr,Adshead:2019igv,Figueroa:2019jsi,Antusch:2020iyq,Antusch:2021aiw} for recent developments and Refs.~\cite{Allahverdi:2010xz,Amin:2014eta} for reviews. At the other extreme lie scenarios in which the only mediator between the inflationary and the reheating sector is gravity, see {\it e.g.}~\cite{Ford:1986sy,Spokoiny:1993kt,Bassett:1997az,Tsujikawa:1999jh,Tsujikawa:1999iv,Tsujikawa:1999me,DeCross:2015uza,DeCross:2016fdz,DeCross:2016cbs,Figueroa:2016dsc,Ema:2016dny,Dimopoulos:2018wfg,Nguyen:2019kbm,Opferkuch:2019zbd,Crespo:2019src,Crespo:2019mmh,vandeVis:2020qcp,Bettoni:2021zhq,Laverda:2023uqv,Bettoni:2018pbl,Bettoni:2019dcw}. In this work we focus on the latter type of scenarios by considering a variation of the {\it gravitational reheating} mechanism originally put forward in Ref.~\cite{Ford:1986sy}. Specifically, we consider a spectator scalar field $\chi$ to be non-minimally coupled to gravity, with an interaction of the form $\xi \chi^2 R$, where $R$ is the Ricci curvature scalar and $\xi$ is a dimensionless coupling. We will refer to $\chi$ from now on as the non-minimally coupled (NMC) scalar field. We also consider an inflaton potential that gives raise to a {\it stiff} equation of state $1/3 < w \leq 1$ after inflation, sustaining a {\it kination} dominated (KD) era. As the name suggests, this is an epoch where the kinetic energy of the inflaton dominates over its potential energy, leading to the inflaton's energy density redshifting faster than radiation~\cite{Turner:1983he,Ford:1986sy}. This effect allows gravitationally produced quanta of $\chi$, whose energy scales as radiation, to come to dominate. In particular, the energy stored in $\chi$, albeit initially tiny compared to the inflaton energy, eventually becomes the dominant energy component in the universe as the inflaton energy redshifts away, ending the KD period. This idea is well exemplified, for example, in {\it Quintessential inflation} scenarios~\cite{Peebles:1998qn,Peloso:1999dm,Huey:2001ae,Majumdar:2001mm,Dimopoulos:2001ix,Wetterich:2013jsa,Wetterich:2014gaa,Hossain:2014xha,Rubio:2017gty}.

The gravitational reheating mechanism as originally formulated in e.g.~\cite{Ford:1986sy,Spokoiny:1993kt,Peebles:1998qn}, has been shown, however, to be inconsistent~\cite{Figueroa:2018twl}. Namely, the (quasi-)scale invariant gravitational wave (GW) spectrum produced during inflation develops a high-frequency blue-tilt due to the fact that radiation modes are {\it blueshifted} during KD~\cite{Giovannini:2009kg, Boyle:2007zx,Figueroa:2018twl,Figueroa:2019paj,Opferkuch:2019zbd,Bernal:2020ywq}. This leads to an excessive amount of energy stored in the resulting GW background, violating cosmic microwave background (CMB) and BBN constraints on the amount of non-standard radiation allowed in the universe~\cite{Caprini:2018mtu,Clarke:2020bil}. While a large number of degrees of freedom can in principle alleviate the problem~\cite{Figueroa:2018twl}, a more theoretically appealing solution can be easily envisaged. In particular, the key observation is to note that the Ricci scalar $R = 3(1-3w)H^2$ flips sign when transitioning from an equation of state $w\approx -1$ during inflation, to a stiff equation of state $w > 1/3$ during KD. A NMC scalar spectator field will in this case acquire an effective tachyonic mass and experience exponential growth during KD~\cite{Figueroa:2016dsc}, rather than behaving as radiation, as originally assumed in e.g.~Refs.~\cite{Ford:1986sy,Spokoiny:1993kt,Peebles:1998qn}. Furthermore, if the field is also self-interacting, its energy will grow only until the self-interactions eventually screen the tachyonic mass. These observations were first put forth in Ref.~\cite{Figueroa:2016dsc}, where the Standard Model (SM) Higgs was considered as a NMC spectator field. In this scenario, the Higgs experiences tachyonic growth during the early stages of KD, but the growth eventually ceases due to the self-interactions of the Higgs (assuming positivity of the Higgs quartic). At this point, the Higgs oscillates and decays into SM particles~\cite{Figueroa:2015rqa,Enqvist:2015sua}, producing a GW background at high frequencies~\cite{Figueroa:2014aya,Figueroa:2016ojl}, and leading to a universe dominated by SM relativistic degrees of freedom. The same mechanism was later studied in more detail in~\cite{Opferkuch:2019zbd}, generalizing the case to arbitrary scalar fields non-minimally coupled to gravity, see also~\cite{Dimopoulos:2018wfg,Bettoni:2021zhq,Laverda:2023uqv}. In~\cite{Opferkuch:2019zbd}, the mechanism was coined as {\it Ricci reheating} and we stick to that nomenclature.
\begin{figure}
	\includegraphics[width = 0.6\textwidth]{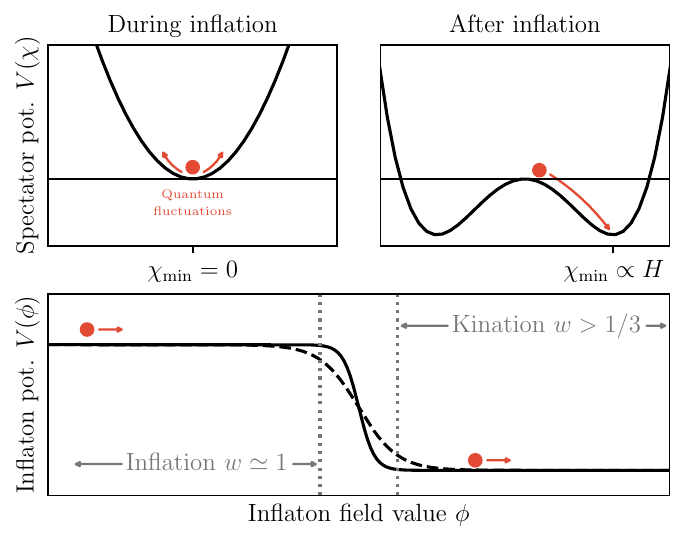}
	\caption{Dependence of the spectator field's effective potential $V_\text{eff}(\chi) \equiv V(\chi) + \frac{1}{2} \xi R \chi^2$ on the dynamics of the inflaton field $\phi$. For illustration purposes we have considered a quartic potential for $V(\chi)$. In the lower panel we show the inflationary potential for two values controlling the transitions speed namely, $\beta = 5$ (2) solid (dashed).}
	\label{fig:sketch}
\end{figure}

A simple example of Ricci reheating can be realized by considering a step-like inflationary potential with the lower region tuned to a comparatively small vacuum energy. KD is achieved when the inflaton field transitions from the higher plateau onto the lower one, see bottom panel of~\cref{fig:sketch}. Once the Universe enters into KD, the inflaton energy density redshifts faster than radiation. In this work, we choose a suitable potential that enables such features, and follow the initial linear dynamics and subsequent non-linear evolution. We consider a realistic transition of finite duration from quasi-de Sitter inflation to KD, which will be referred to, for short, as the qdS-KD transition. Our initial conditions are set deep inside inflation when the relevant fluctuations of the NMC scalar are sub-Hubble. As the qdS-KD transition proceeds, for certain modes there is a change of sign of the effective mass-squared, which flips from positive to negative, signifying the onset of a tachyonic instability. As long as the evolution of the $\chi$ fluctuations is linear, we follow independently the tachyonic growth of the individual modes on top of a time dependent background. The tachyonic instability leads to exponential growth of the NMC field's energy density. This growth eventually ends due to a backreaction of the NMC field itself on the dynamics. In particular, if self-interactions of $\chi$ are negligible, the NMC field eventually comes to dominate the energy budget, driving the Ricci curvature from negative to zero, and hence switching off the instability. If, on the contrary, sufficiently strong self-interactions of $\chi$ are present, these eventually screen the tachyonic mass, and as a result the instability is also switched off. We will refer to the first regularization effect as {\it gravitational backreaction}, and to the second one as {\it self-interaction backreaction}. In order to properly capture the non-linear dynamics that characterize (either) backreaction and the following evolution, we introduce an excited spectrum of $\chi$ on a lattice during the linear regime, just before the onset of any backreaction. We then follow the evolution of the system from that moment onward using classical lattice simulations. For this we use the public code \CL~\cite{Figueroa:2021yhd,Figueroa:2020rrl,Figueroa:2023xmq} and the lattice formulation for NMC scalar field dynamics introduced in~\cite{Figueroa:2021iwm}. In this way, we simultaneously simulate the coupled dynamics of the inflaton and the NMC field, allowing both to fluctuate, and capturing the evolution when the dynamics become fully non-linear, either due to self-interactions, or, as a novelty in this work, throughout the NMC field contribution to the energy budget of the universe. 

The present work is a natural follow-up and closure of Refs.~\cite{Figueroa:2016dsc} and~\cite{Opferkuch:2019zbd}. In the original work, Ref.~\cite{Figueroa:2016dsc} solved the initial dynamics of the NMC field by following the zero mode during the tachyonic phase, while neglecting the expansion of the Universe during the tachyonic phase. Ref.~\cite{Opferkuch:2019zbd} described the dynamics in a more realistic manner, while still following only the zero mode. Contrary to~\cite{Figueroa:2016dsc}, Ref.~\cite{Opferkuch:2019zbd}  considered all necessary terms contributing to the energy density of the NMC field, and correctly incorporated the expansion of the universe throughout all the dynamics. Furthermore, Refs.~\cite{Bettoni:2021zhq,Laverda:2023uqv} have recently studied the non-linear dynamics of the NMC field in the presence of self-interactions by means of classical lattice simulations, though maintaining the inflaton sector as homogeneous, modeling the qdS-KD transition as {\it instantaneous}, and assuming $w = 1$ as exact during KD. 

Our present work extends the use of lattice simulations in Ricci Reheating, finally solving the general problem. In particular, $i)$ we model the qdS-KD transition with a realistic inflaton potential, with parameters that control the transition while satisfying all CMB constraints, and $ii)$ we allow both fields, inflaton ($\phi$) and NMC ($\chi$), to fluctuate, and to potentially dominate the energy budget of the system. We follow all relevant excited modes during the non-linear dynamics in a realistic set-up, including backreaction effects arising from the NMC field, either via its own self-interactions (as considered in previous works), or via its contribution to the energy budget in the universe (as a new consideration in this work). We present a realistic and precise characterization of the e-folding time ($N_{\rm rh}$) and energy/temperature scale ($E_{\rm rh}/T_{\rm rh}$) that characterize Ricci Reheating, assuming NMC self-interactions of the form $V(\chi) = \frac{\lambda}{4}\chi^4$, and including the case when $\lambda \to 0$. While the exact numbers are associated to the chosen family of parametric inflaton potentials, the inflationary enery scale and the duration of the qdS-KD transition in our modeling are tuneable, and hence our results can be expected to be representative for generic situations. 

A particularly interesting application of Ricci Reheating is to consider the SM Higgs as the NMC field. This was, after all, the original motivation of this mechanism in~\cite{Figueroa:2016dsc}. While the latter reference concluded that it is possible to reheat the Universe at high temperatures without probing the unstable region of the SM Higgs~\cite{Degrassi:2012ry,Bezrukov:2012sa}, Ref.~\cite{Opferkuch:2019zbd} concluded however that, once the Hubble rate is allowed to evolve during the tachyonic growth of $\chi$, only reheating temperatures below $T_{\rm rh} \lesssim 1$ GeV are allowed. Ref.~\cite{Laverda:2024qjt} has recently revisited the problem and claimed that if an improved running of the SM Higgs self-coupling is considered, the reheating temperatures can be significantly higher, as $T_{\rm rh} \lesssim 10^{5}$ GeV. While the eventual goal is a reexamination of this possibility in a realistic set-up like ours, we postpone addressing this question for a future study, where the results presented here will serve as an important input. 

This paper is divided as follows. In \cref{sec:model}, we introduce the model, present the dynamical equations that we need to solve, and discuss the initial conditions required to start the lattice simulations. In \cref{sec:results}, we present our results for both gravitational and self-interaction backreaction scenarios, including comparisons to the literature where appropriate. We conclude and summarize our findings in \cref{sec:summary}. 


\section{Model: setup, initial conditions, and dynamics}
\label{sec:model}
\noindent
We consider a model involving a minimally coupled inflaton field $\phi$ and a non-minimally coupled (NMC) spectator scalar field, $\chi$, which we will refer to as the {\it reheaton}. The complete action of the model is
\begin{align}
	\mathcal{S}=\int \diff^4x \sqrt{-g} \left[ \frac{m_p^2}{2} R + \mathscr{L}_{\chi} + \mathscr{L}_\text{inf} \right] \,,
\label{eq:act}
\end{align}
where $R$ is the Ricci scalar, $m_p = 2.435 \times 10^{18}$ GeV is the reduced Planck scale, and we have defined
\begin{align}
\label{eq:nmc-l}
	\mathscr{L}_{\chi} &= -\frac{1}{2} g^{\mu\nu}\partial_{\mu}\chi\partial_{\nu}\chi -  \frac{1}{2}\xi R \chi^{2}- V(\chi)  \,, \\
	\mathscr{L}_\text{inf} &= -\frac{1}{2} g^{\mu\nu}\partial_{\mu}\phi\partial_{\nu}\phi - V_\text{inf}(\phi)\,,
\label{eq:inf-l}
\end{align}
with $\xi \neq 0$ a dimensionless real parameter that quantifies the strength of reheaton's non-minimal coupling to gravity.

Crucial to the Ricci reheating mechanism is an inflaton potential that allows for a transition from a quasi-de Sitter inflationary phase to a kination-dominated (KD) universe. A simple example that exhibits slow-roll inflation with a rapid transition to kination thereafter is a double plateau-like potential, as considered e.g.~in~\cite{Spokoiny:1993kt}. We show a sketch of such a potential in \cref{fig:sketch}, where one can see that the transition from the inflationary phase ($w \simeq -1$) to the kination phase ($w >  1/3$) causes the effective mass-squared of the NMC field $m_{\chi}^{2} \simeq 3(1-3w)H^2$ to flip sign and become tachyonic. This tachyonic mass results in an exponential growth of the reheaton field amplitude until it is regulated by either self-interactions in the potential $V(\chi)$, or by conservation of energy occurring if the reheaton field comes to dominate the energy budget in the universe. In the latter case, the non-minimally coupled reheaton sources the expansion of the universe, which we handle via the techniques developed in Ref.~\cite{Figueroa:2021iwm}. We thus consider the following possibilities:
\begin{enumerate}
\item {\bf Self-interaction backreaction}: This corresponds to the case of a quartic potential, $V(\chi) = \frac{1}{4} \lambda \chi^4$. The tachyonic growth of $\chi$ will be regulated by a screening of its negative gravitational mass, $m_{\rm grav}^2 = \xi R < 0$, by a positive effective mass induced through self-interactions $m_{\rm int}^2 = \lambda \langle\chi^2\rangle > 0$, once the amplitude of $\langle\chi^2\rangle$ has grown above a critical threshold.

\item {\bf Gravitational backreaction}: This corresponds to the case of a free theory, i.e. $V(\chi) = 0$, or alternatively, to $\lambda$ sufficiently small.  The tachyonic growth of $\chi$ will be regulated once $\chi$ dominates the energy budget of the universe, as $R$ is driven to zero in those moments, shutting off the tachyonic instability.
\end{enumerate}
For concreteness, we choose the following double plateau-like potential for the inflaton
\begin{align}\label{eq:Vinf}
	V_\text{inf}(\phi)&= \frac{V_*}{2} \left[1-\tanh\left(\frac{\beta \phi}{m_p} \right)\right]\,,
\end{align}
where $\beta$ is a dimensionless constant that controls speed of the qdS-KD transition, and $V_*$ is a dimension-4 constant determining the scale of inflation. We identify the inflationary Hubble rate when CMB scales left the Hubble radius as $H_*^2 \simeq V_* / 3 m_p^2$, with this identity holding to better than $\mathcal{O}(0.1)\%$ in the cases considered in this work. We focus on $\beta > 1$, as this leads to a somewhat fast transition compared to the Hubble time $H^{-1}$ at the end of inflation. We note that $\beta \leq 1$ does not actually work for our purposes, as the inflaton does not reach KD due to Hubble friction. Without loss of generality, we chose the negative branch of the potential (${\phi} < 0$) to sustain the inflationary phase, and define the number of e-folds such that $N=0$ marks the end of inflation. 

\begin{figure}
        \includegraphics[width = 0.495\textwidth]{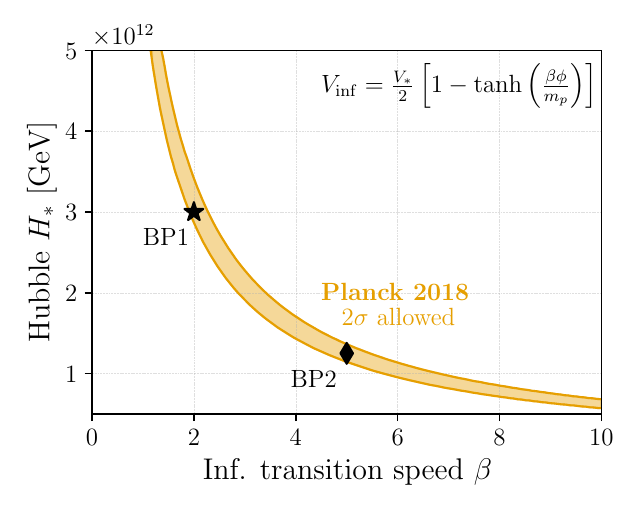}
	\caption{Allowed region of parameter space ({\color{cborange}\bf shaded orange}) that satisfies all inflation constraints from \cref{eq:AsNsR,eq:CMBconsA,eq:CMBconsNs,eq:CMBconsR}. Additionally the two benchmark points we consider further are shown.}
	\label{fig:infPotConstraints}
\end{figure}

Using current CMB constraints on scalar and tensor perturbations, 
we can constrain the parameters $\lbrace V_*, \beta \rbrace$ that characterize the potential in \cref{eq:Vinf}, by confronting its inflationary predictions 
against CMB observations. 
In order to do this, we calculate the standard (potential) {\it slow-roll} parameters $\epsilon_V \equiv (m_p^2/2)\left(V_{\rm inf}'/V_{\rm inf}\right)^2$ and $\eta_V \equiv m_p^2\left(V_{\rm inf}''/V_{\rm inf}\right)$, and then use the slow-roll machinery to express the spectral amplitude and index of scalar perturbations, $A_s, n_s$, as well as the tensor-to-scalar ratio, $r \equiv A_t/A_s$, as a function of $\epsilon_V, \eta_V$~\cite{Baumann:2009ds}
\begin{align}\label{eq:AsNsR}
A_s \simeq \frac{1}{24\pi^2\epsilon_{V_k}}\frac{V_{\rm inf}({\phi}_{k})}{m_p^4}\,,~~~~ n_s - 1 \simeq 2\eta_{V_k} - 6\epsilon_{Vk}\,,~~~~ r_k \simeq 16\epsilon_{V_k}\,,
\end{align} 
where a sub-index $_k$ indicates the time when the scale $k$ crossed the Hubble radius $- N_k$ e-folds before the end of inflation. In particular, we use the latest CMB constraints~\cite{Planck:2018jri,BICEP:2021xfz,Tristram:2021tvh} which, at the pivot scale $k_{\rm CMB}=0.05$~Mpc$^{-1}$,~read
\begin{align}\label{eq:CMBconsA}
    A_s &= 2.099^{+0.296}_{-0.292}\cdot 10^{-9}~(68\% ~{\rm CL})\,,~~~
    \\
    \label{eq:CMBconsNs}
    n_s &= 0.9649 \pm 0.0042~(68\% ~{\rm CL})\,,\\
    \label{eq:CMBconsR}
    r &< 0.032 ~(95\% ~{\rm CL})\,. 
\end{align}
The orange region in \cref{fig:infPotConstraints} shows the area in the $\beta$-$H_*$ plane satisfying the above constraints. This region is determined assuming that CMB scales left the Hubble radius between 50-60 e-foldings before the end 
of inflation. Given the plateau shape of the branch of potential~(\ref{eq:Vinf}) that sustains inflation, Hubble rates $H_* < 10^{12}$ GeV can only be compatible with CMB constraints for $\beta \gg 1$, as $H_{*} \propto \beta^{-1}$. In this case, the dynamics following inflation become insensitive to the transition, which would be perceived as {\it instantaneous} compared to the Hubble scale. At the same time, the colored region in \cref{fig:infPotConstraints} indicates that the energy scale in our model cannot be much smaller than the upper bound set by the constraint~(\ref{eq:CMBconsR}), which requires $H_* \leq H_{\rm max} \simeq 4.43 \times 10^{13}$ GeV. Furthermore, we note that in the case of slow transitions ($\beta \lesssim 1$), the system fails to transition to KD after inflation, as the inflaton's kinetic energy never dominates sufficiently over its potential. In light of all the above discussion, we select benchmark points (BP's) for our analysis with Hubble scales smaller than the CMB upper bound, say $H_* \sim \mathcal{O}(0.1)H_{\rm max} \sim 10^{12}$ GeV, such that $1 \lesssim \beta \lesssim 10$. This guarantees a somewhat fast transition compared to the Hubble scale at the end of inflation, but also lies sufficiently far away from the instantaneous limit ($\beta \gtrsim 10$) that has already been studied in previous work. We choose BP's with different values of $\beta$, so that we can compare the dependence of the resulting dynamics on different finite transition rates: 
\begin{itemize}
	\item BP1 ($\bigstar$): Fast transition, $\beta = 2$, $H_* = 3\times 10^{12}$ GeV,
	\item BP2 ($\blacklozenge$): Very fast transition, $\beta = 5$, $H_* = 1.25\times 10^{12}$ GeV.
\end{itemize}

\subsection{Equations of motion}

Having discussed the model we now move on to introduce the equations of motion that describe the dynamics of the system. We assume a flat Friedmann-Lema\^itre-Robertson-Walker (FLRW) background
\begin{equation}
ds^{2} =-a(\eta)^{2\alpha}d\eta^{2}  + a(\eta)^{2}  \delta_{ij}dx^{i}dx^{j}  \,,
\label{eq:rt_metric}
\end{equation}
where $d\eta$ is an ``$\alpha$-time" variable related to cosmic time as $dt = a(\eta)^{\alpha} d\eta$. For this metric, the Friedmann equation for the scale factor $a(\eta)$ sourced by both minimally coupled ($\phi$) and non-minimally coupled ($\chi$) sectors, together with the equations of motion for $\phi$ and $\chi$, have been derived in~\cite{Figueroa:2021iwm}, and read
\begin{align}
\phi'' +(3-\alpha) \frac{a'}{a}\phi' - a^{-2(1-\alpha)}\nabla^2\phi   &= - a^{2\alpha}  V'_\mathrm{inf}(\phi) 
\label{eq:inflatonEOM}
\\
\chi'' +(3-\alpha) \frac{a'}{a}\chi' - a^{-2(1-\alpha)}\nabla^2\chi  &= - a^{2\alpha} \Big[\xi R \chi + V'(\chi)\Big] \label{eq:NMCeom} \,,\\
\label{eq:piadot}
     \frac{a''}{a} + (1-\alpha)\left(\frac{a'}{a}\right)^2 &= \frac{a^{2\alpha}}{6} R \; ,
\end{align}
where the Ricci scalar is given by
\begin{align}\label{eq:Ricci}
    R &= \frac{F(\chi)}{m_p^{2}}\Big(\left(1-6\xi \right) \langle\partial^{\mu}\chi\partial_{\mu}\chi\rangle  + 4 \langle V(\chi)\rangle- 6\xi\langle \chi V'(\chi)\rangle 
    + \langle\partial^{\mu}\phi\partial_{\mu}\phi\rangle  + 4 \langle V_{\rm inf}(\phi)\rangle
     \Big)\,,\\
\label{eq:FChi}
\hspace{-2cm}{\rm with}\hspace{2cm}&   \hspace{-0.75cm}F(\chi) = {1\over 1 + \left(6\xi -1\right)\xi \langle\chi^2\rangle /m_p^2 } \,. 
\end{align}
The specific integration scheme we implement to solve this coupled system of non-linear differential equations~(\ref{eq:inflatonEOM})-(\ref{eq:piadot}) on a lattice, is outlined in detail in Section II of Ref.~\cite{Figueroa:2021iwm}. Here we simply note that the Hubble rate 
\begin{align}\label{eq:HubbleBothFlds}
    \mathcal{H}^2 \equiv& \left(\frac{a'}{a}\right)^2 = \frac{a^{2\alpha}}{3 m_p^2}\Big(\langle \rho_\phi \rangle+ \langle \rho_\chi \rangle \Big) \:,
\end{align}
can be used as a constraint equation to check the consistency of the dynamics, as it is sourced from the homogeneous energy density of both the inflaton and the NMC fields,
\begin{eqnarray}\label{eq:InflatonEnergy}
    \langle \rho_\phi \rangle &=& \frac{1}{2a^{2\alpha}}\left\langle\phi'^2\right\rangle + \frac{1}{2a^{2}}\left\langle(\nabla \phi)^2\right\rangle + \left\langle V_{\text{inf}}(\phi) \right\rangle,  \\
    \label{eq:NMCenergy}
    \langle \rho_\chi \rangle &=& \frac{1}{2a^{2\alpha}}\left\langle\chi'^2\right\rangle + \frac{1}{2a^{2}}\left\langle(\nabla \chi)^2\right\rangle + \left\langle V(\chi) \right\rangle + \frac{3 \xi}{a^{2\alpha}} \mathcal{H}^2\left\langle \chi^2 \right\rangle + \frac{6\xi}{a^{2\alpha}} \mathcal{H} \left\langle \chi \chi'\right\rangle,
\end{eqnarray}
where $\langle ... \rangle$ represents volume averaging.

To account for the full non-linear dynamics, we solve the set of coupled partial differential equations~(\ref{eq:inflatonEOM})-(\ref{eq:piadot}) using the $\mathcal{C}osmo\mathcal{L}attice$ public package~\cite{Figueroa:2020rrl, Figueroa:2021yhd, Figueroa:2023xmq} and its module for NMC fields based on~\cite{Figueroa:2021iwm}. In particular, space is regularized by introducing a cubic lattice of size $(N\cdot \dd x)^3$, with $N$ an integer and $\dd x$ the lattice spacing (in comoving coordinates). The equations are discretized using a second order finite-difference scheme and they are evolved in time by a second order Runge-Kutta method, see Ref.~\cite{Figueroa:2021iwm} for details. For the simulations presented here we have utilized lattices with $N=400-600$ points per spatial dimension and fine time-step of $H_*\delta t = \num{2.5E-3}$. We note that such a small step is necessary to resolve the most ultraviolet (UV) scales of the NMC field spectra at late times in the presence of self-interactions. During evolution of the system we check that Eq.~(\ref{eq:HubbleBothFlds}) is consistent, plugging in the solution for $\lbrace a(\eta), \langle \rho_\phi \rangle, \langle \rho_\chi \rangle \rbrace$ obtained from solving Eqs.~(\ref{eq:inflatonEOM})-(\ref{eq:piadot}). In practice, in our lattice runs we always verify Eq.~(\ref{eq:HubbleBothFlds}) to better than $0.1\%$ accuracy for all subsequent results.
 
\subsection{Initial conditions and lattice computation}
\label{subsec:power_spec}

Regarding the initial conditions of our problem, the inflaton field is set in a standard manner consistent with random initial fluctuations. We assume the inflaton is dominantly homogeneous and thus it evolves according to the homogeneous Klein-Gordon equation. Fluctuations are then added on top as random realizations of Fourier modes, following a Gaussian distribution with zero mean and variance given by a spectrum mimicking quantum vacuum fluctuations, see either of Refs.~\cite{Figueroa:2020rrl,Figueroa:2021yhd} for details on this standardized procedure. The initial spectrum of the field $\chi$ requires however extra care, due to the presence of the non-minimal coupling. Fixing for this discussion $\alpha=1$, namely conformal time, we observe that
the kinetic term for $\chi$ can be made canonical via the field redefinition $\chi \rightarrow \sigma/a$ which, after removing total derivatives, leads to the following action for $\sigma$ (where we neglect for the moment self-interactions of the NMC field), 
 \begin{equation}
\mathcal{S}_\sigma = \frac{1}{2} \int \diff\tau \diff^{3}x \left[(\sigma')^{2} - (\nabla\sigma)^{2} - a^{2}\left(\xi - \frac{1}{6}\right)R\,\sigma^{2} \right] \,,
\label{eq:Ract}
\end{equation}
where we have used that in a FLRW background, $a''/a^3 = R/6$. Quantization of $\sigma$ then proceeds as in Minkowski space~\cite{Birrell:1982ix}
\begin{equation}
\hat{\sigma} (\tau, x) = \int \frac{d^{3} k}{(2\pi)^{3}} \left[ \sigma_{k}(\tau) \hat{a}_{k} e^{ikx} + \sigma_{k}^{*} (\tau) a^{\dagger}_{k}e^{-ikx}\right] \,,
\end{equation}
where the creation and annihilation operators satisfy the usual algebra $[\hat{a}_{k},\hat{a}_{k'}^{\dagger}] = (2\pi)^{3} \delta(k-k')$ and the mode fluctuations are normalized as $\sigma_{k}\sigma_{k}'^{*} -\sigma_{k}' \sigma_{k}^{*} = i$. The mode functions $\sigma_{k}(\tau)$ satisfy the following equation of motion
 \begin{equation}
\sigma''_{k} + \left[k^{2} + a^{2}\left(\xi - \frac{1}{6}\right) R \right] \, \sigma_{k} = 0 \,.
\label{eq:modeEqConf}
\end{equation}
The coincident, equal-time two point function can be computed as
\begin{equation}\label{eq:sigma2}
\langle \sigma^{2} \rangle = \langle 0| \hat{\sigma}(\tau, {\bf x})\hat{\sigma} (\tau, {\bf x}) |0 \rangle = \int \frac{dk}{k} \Delta_{\sigma}(k,\tau) \,,
\end{equation}
where we have introduced a power spectrum defined as
\begin{align}\label{eq:PS}
\Delta_{\sigma}(k,\tau) = \frac{k^{3}}{2\pi^{2}} \mathcal{P}_{\sigma}(k,\tau)\,,~~~~~{\rm with}~~\langle \sigma_{{\bf k}}\sigma_{{\bf k}'}^* \rangle = (2\pi)^3 \mathcal{P}_{\sigma}(k,\tau) \delta({\bf k}-{\bf k}')\,.
\end{align} 
\begin{figure}
	\includegraphics[width=0.495\textwidth,trim={0.4cm 0.4cm 0.4cm 0.4cm}, clip]{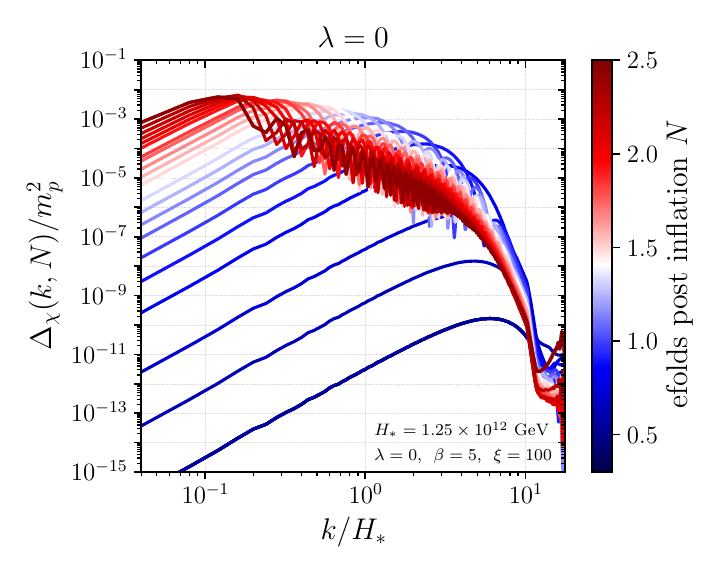}
 	\includegraphics[width=0.495\textwidth,trim={0.4cm 0.4cm 0.4cm 0.4cm}, clip]{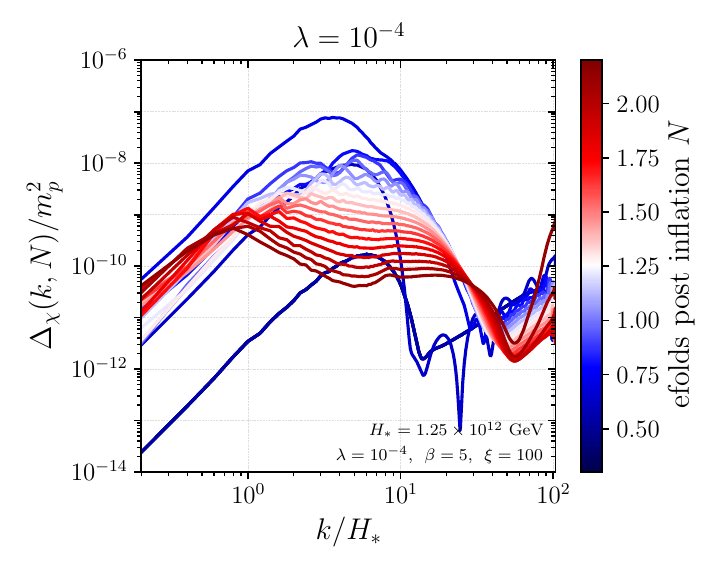}
	\caption{Power spectra of $\chi$ during the initial tachyonic growth regime and throughout the backreaction process, when $R < 0$ in KD for a free-field (left) and the case with a quartic $\lambda = \num{e-4}$ (right). The spectrum at the earliest time ($N=0.3$) determined from solving \cref{eq:modeEqConf}, is the initial condition for the lattice simulation.}
	\label{fig:power-spetra}
\end{figure}
\noindent The power spectrum for $\chi$ is simply given by $\Delta_{\chi}(k,\tau) = a^{-2} \Delta_{\sigma}(k,\tau)$. As shown in Fig.~\ref{fig:power-spetra}, such power spectra typically exhibit a large excitation within a finite range of momenta. In order to understand such a {\it bump} in the spectra, 
we note that during the period following immediately after inflation, $R$ is only sourced by the homogeneous inflaton, which initially dominates the energy budget. Introducing $K(\phi) \equiv \frac{1}{2 a^{2\alpha} } {\phi'}^2$, $G(\phi) \equiv \frac{1}{2 a^2}  (\vec\nabla \phi)^2$ and $V_{\rm inf}(\phi)$ as the inflaton's kinetic, gradient and potential energy densities, we observe that the equation of state of the universe becomes, as the inflaton enters into KD ($K > 2V$), 
\begin{align}\label{eq:w-def}
w \equiv {\left\langle K(\phi) - V_{\rm inf}(\phi)\right\rangle \over \left\langle K(\phi) + V_{\rm inf}(\phi)\right\rangle} > 1/3\,,~~~ ({\rm KD})
\end{align}
gradually approaching $w \rightarrow 1$. The transition to unity occurs faster for larger $\beta$, namely when the qdS-KD transition is fast compared to the Hubble scale. As a result the Ricci scalar become negative, as
\begin{align}\label{eq:Rinf}
R = {2\over m_p^2}\Big\langle2V_{\rm inf}(\phi)-K(\phi)\Big\rangle < 0\,,~~~ ({\rm KD})\,.
\end{align}
We note that $\langle ... \rangle$ in Eqs.~(\ref{eq:w-def}) and (\ref{eq:Rinf}) represents volume averaging, whereas in \cref{eq:sigma2} and \cref{eq:PS} it represents quantum-mechanical ensemble averaging.

At times when the gravitational backreaction of $\chi$ can be ignored, and its self-interactions can be neglected, \cref{eq:modeEqConf} represents a linear equation for $\sigma_k$, for which $R < 0$ during KD. In this linear regime, \cref{eq:modeEqConf} indicates that, as long as $\xi > 1/6$, infrared modes below a certain cutoff $k \lesssim k_*$, experience a {\it tachyonic} instability leading to exponential growth as $\sigma_k \propto e^{|\omega_k|\tau}$. From \cref{eq:modeEqConf}, we see that $\omega_k^2 \equiv (k^{2} + a^{2}(\xi - \frac{1}{6}) R ) < 0$, and hence we deduce that the tachyonic cutoff scale is
\begin{equation}\label{eq:UVcutoff}
k_* = \sqrt{(\xi-1/6) |R| a^2} \:.
\end{equation}
Using the universal relation in a FLRW background~\cite{Figueroa:2021iwm}, $R = 3(1-3w)H^2$, with $H^2$ the Hubble rate, we also conclude that $R < 0$ for $w > 1/3$. As the sound speed of a fluid must not exceed the speed of light, $c_{\text{s}}^2=\partial p/\partial \rho \leq 1$, the equation of state during KD is also bounded as $w \leq 1$. Consequently, the most negative value of $R$ is also bounded from below as $R\geq -6H^2$. This implies that if $\xi \gg 1$, sub-horizon modes of the NMC field will be excited, up to a maximum scale $k_* \simeq \sqrt{6\xi}aH$, as long as KD with $w \simeq 1$ is sustained. 

It is precisely the tachyonic instability during the linear regime, that actually explains the form of the spectra shown in Fig.~\ref{fig:power-spetra}. The initial single bump shape is simply a consequence of the fact that only infrared modes up to a cutoff scale $k \lesssim k_*$, are excited during the linear regime. Above such cutoff, i.e.~for $k \gtrsim k_*$, modes remain in vacuum. These vacuum fluctuations actually give rise to UV divergences in the power spectrum, which technically need to be regularized~\cite{Ferreiro:2022ibf,Ferreiro:2023uvr,Pla:2024xsv}. However, in practice on the lattice, the amplitude of the excited bump is much larger than the UV tail due to vacuum fluctuations. This is ensured thanks to the fact that the natural UV cutoff on the lattice, $k_{\rm UV} = \pi/\delta x$, is typically of the same order as (though slightly larger than) $k_*$. Therefore, as it is customary in lattice simulations~\cite{Figueroa:2020rrl}, we simply allow vacuum fluctuations to remain present in the simulation, as they do not affect the dynamics of the excited modes. In practice we ensure the above by choosing to start the simulation some time after inflation such that a clearly defined bump has formed, initializing the lattice simulation with such a peaked spectrum, {\it c.f.} the earliest time spectra in \cref{fig:power-spetra}. From this point onwards, we run the lattice version of the equations of motion previously introduced, Eqs.~(\ref{eq:inflatonEOM})-(\ref{eq:piadot}), which will capture the dynamics of the two-field system, even when non-linearities develop later on. We present our results on the non-linear dynamics in the following section.

\section{Results}\label{sec:results}

As explained above, the tachyonic instability during the linear regime leads to exponential growth of the NMC field amplitude, and hence of its energy density. However, due to energy conservation, such growth cannot be sustained indefinitely. Basically, once the NMC energy surpasses a given threshold, the backreaction of $\chi$ cannot be longer ignored, and the dynamics become non-linear from then on. As a result, the tachyonic instability is gradually switched off. The nature of the backreaction, and hence of the non-linear dynamics, depends crucially on whether the NMC field sustains large self-interactions or not. If self-interactions of $\chi$ are negligible, the NMC field eventually comes to dominate the energy budget of the universe, driving $R$ to zero, and thus switching off the instability. If, on the contrary, sufficiently strong self-interactions of the NMC field are present, these eventually screen the tachyonic mass, and as a result the instability is also switched off (even though $\chi$ is still typically energetically very subdominant). We refer to the first regularizing effect as {\it gravitational backreaction}, and to the second one as {\it self-interaction backreaction}. In what follows, we describe both non-linear regimes in light of our lattice results, with the aid of analytic expressions. 

\subsection{Ricci Reheating via gravitational backreaction}
\label{sec:grav-br}
\begin{figure}  
 	\includegraphics[width=0.495\textwidth,trim={0.4cm 0.4cm 0.4cm 0.4cm}, clip]{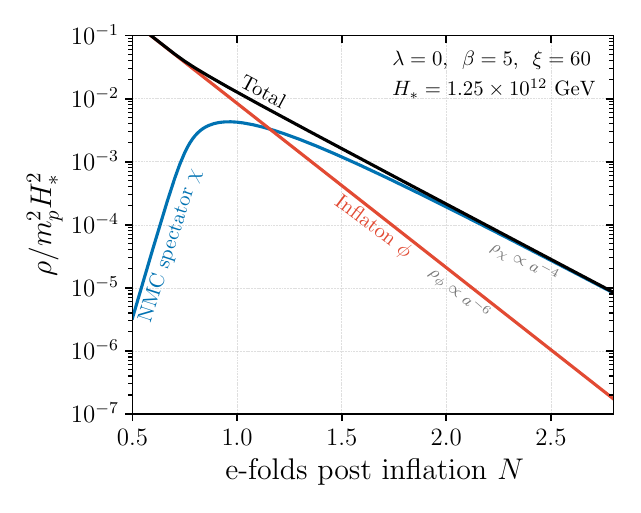}
   \includegraphics[width=0.495\textwidth,trim={0.4cm 0.4cm 0.4cm 0.4cm}, clip]{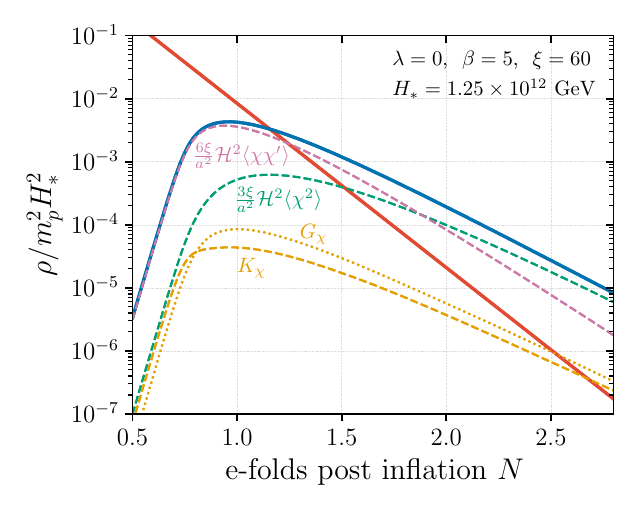}
	\caption{Evolution of the energy densities as a function of e-folds since the end of inflation, for $V(\chi) = 0$. {\bf Left:} Energy density of the spectator field $\chi$ ({\color{cbblue}\bf blue}), the inflaton $\phi$ ({\color{defred}\bf red}) and the total energy density ({\bf black}). Note that $\rho_\chi$ scales as radiation at late times once it comes to dominate over the inflaton. {\bf Right:} Components contributing to the NMC field $\chi$'s energy density. The new components from the non-minimal coupling $\xi$ ({\color{cbpurple}\bf purple} \& {\color{cbgreen}\bf green}) dominate over the standard contributions from the gradients ({\color{cborange}\bf orange dotted}) and kinetic term ({\color{cborange}\bf orange dashed}) defined above \cref{eq:w-def}.} 
	\label{fig:NMC_energy_densities}
\end{figure}

\begin{figure}  \includegraphics[width=0.495\textwidth,trim={0.4cm 0.4cm 0.4cm 0.4cm}, clip]{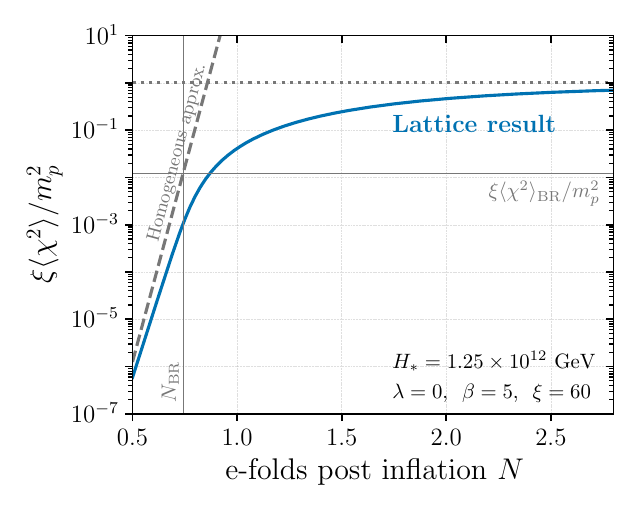} 
	\caption{Growth of $\langle \chi^2 \rangle$ for gravitational backreaction scenario, that is $\lambda = 0$. The full lattice result is shown in {\color{cbblue}\bf blue}, which as argued in the text asymptotes to one.  The result from Ref.~\cite{Opferkuch:2019zbd} is shown as a {\color{defgrey}\bf dashed grey} line where the system is solved assuming $\chi$ is homogeneous, without including gravitational backreaction. In addition the {\color{defgrey}\bf thin grey} vertical and horizontal lines are the approximate analytic results, see \cref{eq:chi2_analytic,eq:BR_expressions}.}
	\label{fig:NMC_field_behavior}
\end{figure}

We first consider the case where the reheaton field $\chi$ has no potential\footnote{As we will see in Sect.~\ref{subsec:RRviaSelfInteraction}, this is simply equivalent to choosing very weak interactions with self-coupling $\lambda \lesssim 10^{-10}$.}, that is $V(\chi) = 0$. As can be seen in Fig.~\ref{fig:NMC_energy_densities}, the NMC field undergoes tachyonic growth until its energy density becomes comparable with that of the inflaton. The growth then ceases since the reheaton field must satisfy the Einstein equations as the inflaton energy continues to redshift away roughly as $a^{-6}$. The backreaction is such that $\chi$ asymptotically approaches a constant, as shown in \cref{fig:NMC_field_behavior}. By combining \cref{eq:HubbleBothFlds,eq:Ricci} and using the universal relation $R = 3(1-3w)H^2$~\cite{Figueroa:2021iwm}, one can show that solutions static in $\chi$ must satisfy 
\begin{equation}
(6\xi -1) F(\chi) \left(1-\frac{\xi \langle \chi^2 \rangle}{m_p^2} \right) \frac{\xi \langle \chi^2 \rangle}{m_p^2} = 0 \,.
\end{equation}
For a non-vanishing and non-conformal 
parameter $\xi \neq 0$ and $\xi \neq 1/6$, the only solution where $\langle \chi^2 \rangle \neq 0$ is $\langle \chi^2 \rangle = m_p^2/\xi$, since $F(\chi)$ never vanishes, {\it c.f.~}\cref{eq:FChi}. For a static configuration of $\chi$, the Ricci scalar can be written as
\begin{equation}
a^{2\alpha} R = 6\left( \frac{a'}{a}\right)^{2} F(\chi) (1-6\xi)\left(1-\frac{\xi \langle \chi^2 \rangle}{m_p^2} \right) \,,
\end{equation}
so we see that we have $R\rightarrow 0$ as $\langle \chi^2 \rangle \rightarrow m_p^2/\xi$, {\it c.f.~}Fig.~\ref{fig:NMC_field_behavior}. Thus, gravitational backreaction drives $R$ to zero, and as  consequence the reheaton-dominated universe expands as a radiated-dominated universe, essentially due to the lack of any non-gravitational mass scales in the system. This can be seen clearly in the left-hand panel of \cref{fig:NMC_energy_densities}, where the scaling of the energy density at late times behaves as $a^{-4}$.

We can get an approximate analytic handle on the gravitational backreaction by noting that according to the right-hand panel of \cref{fig:NMC_energy_densities}, gradients give a very subdominant contribution to the energy density of $\chi$. Thus, it is a good approximation to assume that $\chi$ initially obeys a homogeneous Klein-Gordon equation, which in real time reads
\begin{equation}
\ddot{\chi} + 3H\dot{\chi} + \xi R \chi \approx 0 \,,
\end{equation}
which, for $w=1$, has a growing solution of the form
\begin{equation}
\chi(t) \approx \frac{\chi_*}{2} (1+\sqrt{2}) \left(\frac{H_*}{H}\right)^{\sqrt{2\xi/3}} \,,
\label{eq:homoExpSoln}
\end{equation}
where $\chi_* = 0.06 H_* / \xi^{1/4}$ according to the initial conditions given in Ref.~\cite{Opferkuch:2019zbd}. Similarly, neglecting gradients, the energy density of $\chi$ reads
\begin{equation}
    \langle \rho_{\chi} \rangle \approx \frac{1}{2a^{2\alpha}}\left\langle\chi'^2\right\rangle  + \frac{3 \xi}{a^{2\alpha}} \mathcal{H}^2\left\langle \chi^2 \right\rangle + \frac{6\xi}{a^{2\alpha}} \mathcal{H} \left\langle \chi \chi'\right\rangle \approx 6\xi (1+\sqrt{6\xi}) \frac{\mathcal{H}^2}{a^{2\alpha}} \left\langle \chi^2 \right\rangle \;,
\end{equation}
\begin{figure}
	\includegraphics[width=0.83\textwidth,trim={0.35cm 0.25cm 0.35cm 0.45cm}, clip]{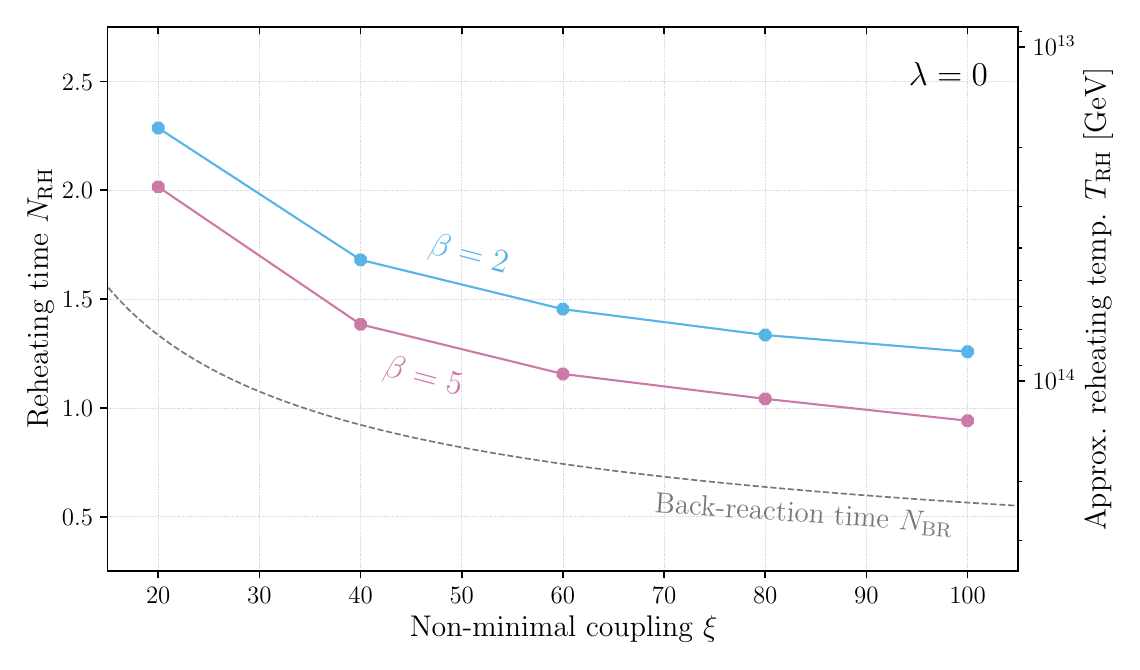}
	\caption{Number of e-folds for the NMC spectator fields' energy density to equal the inflaton energy density (defined as $N_\mathrm{RH}$) for $\beta = 2\,(5)$ {\color{cblblue}\bf light-blue} ({\color{cbpurple}\bf purple}) which parameterize the steepness of the inflationary potential in \cref{eq:Vinf}. The dots indicate the points where a lattice simulation was run, see the text for details. We also show ({\color{defgrey}\bf grey dashed}) the na\"ive time until the field growth backreacts on the Ricci scalar terminating the tachyonic growth of the NMC spectator field, see \cref{eq:BR_expressions}. For illustrating the reheating temperature, we take $H_* = \SI{1.25E+12}{\GeV}$ and the effective relativistic degrees of freedom $g_{\rm RH} = g_\text{SM} + 1 = 107.75$ . The exact reheating temperatures for $\beta = 2$ are given in \cref{tab:grav-backreaction-lattice-results}.}
	\label{fig:RHtime}
\end{figure}
where in the last approximate equality we have used~\cref{eq:homoExpSoln}. At the moment when the inflaton and NMC field energies are the same, they should carry half of the critical density $\rho_{\rm crit} = 3m_p^2 \mathcal{H}^2/a^{2\alpha}$ each. This allows us to solve for $\langle \chi^2 \rangle$ at the time of backreaction
\begin{equation}\label{eq:chi2_analytic}
\frac{\xi \left\langle \chi^2 \right\rangle_{\rm BR} }{m_p^2} \approx \frac{1}{4 (1+\sqrt{6\xi}) } \approx \frac{\xi \chi_*^2 }{m_p^2} \left( \frac{H_*}{H_{\rm BR}}\right)^{2\sqrt{\frac{2\xi}{3}}} \,,
\end{equation}
where the subscript $``{\rm BR}"$ indicates the moment of backreaction. From this we find
\begin{equation}\label{eq:BR_expressions}
N_{\rm BR} = \frac{1}{2\sqrt{6\xi}} \log \left( \frac{\left\langle \chi^2 \right\rangle_{\rm BR}}{\chi_*^2}\right) \,, \hspace{15mm} \frac{\rho_{\rm BR}}{m_p^2 H_*^2} \approx \frac{3}{2}\left( \frac{\left\langle \chi^2 \right\rangle_{\rm BR}}{\chi_*^2}\right)^{-\sqrt{\frac{3}{2\xi}}} \,,
\end{equation}
where $N_{\rm BR}$ is the backreaction time in terms of the number of e-folds and $\rho_{\rm BR}$ is the corresponding energy density. Both of these quantities are shown as thin grey horizontal/vertical lines in \cref{fig:NMC_field_behavior}. We observe that $N_\text{BR}$ is faster compared to the corresponding lattice results, for two main reasons. Firstly, the initial amplitude of $\chi$ is overestimated in Ref.~\cite{Opferkuch:2019zbd} due the assumptions of an instantaneous qdS-KD transition and spatial coherence of the NMC field. Secondly, the backreaction begins to slow down the exponential growth once $\xi \langle \chi^2\rangle/m_p^2$ reaches $\mathcal{O}(0.1\%)$, as evident from \cref{fig:NMC_field_behavior}. 

We define the moment of reheating to be when the reheaton energy actually equals that of the inflaton in our simulation. In~\cref{fig:RHtime}, we show how many e-folds $N_{\rm RH}$ it takes to reach this point as a function of the non-minimal coupling $\xi$, for two different values of the transition parameter $\beta$. One can see that if $\xi \gg 1$ such that the tachyonic growth of the field can overcome the expansion of the universe, this typically occurs in an $O(1)$ number of e-folds, with only a mild dependence on the transition parameter $\beta$. This therefore constitutes a very efficient reheating mechanism if self-interactions of the field can be neglected. For comparison the grey-dashed line shows the results taking the backreaction time to equal to the reheating time, which is what one would get by naively extrapolating the exponential growth phase until the NMC field energy becomes equal to that of the inflaton. It is not surprising that in general this leads to an underestimation of $N_{\rm RH}$ and an overestimation of $T_{\rm RH}$, since it does not take into account the fact that the non-linear dynamics of gravitational backreaction delay the moment of reheating. Nevertheless, the backreaction time is useful since it gives a lower bound on the reheating time and it indicates the onset of non-linearities in the system. We summarize the values entering \cref{fig:RHtime} in \cref{tab:grav-backreaction-lattice-results}.

\begin{table}[]
    \centering
    \begin{tabular}{C{0.5cm}  C{2cm}  C{2cm} C{2cm} C{2.5cm}}  \toprule
           & $\xi$ & $N_\text{RH}$ &  $N_\text{BR}$ &  $T_\text{RH} \text{ [GeV]}$
          \\\midrule 
         \parbox[t]{2mm}{\multirow{5}{*}{\rotatebox[origin=c]{0}{$\beta =2$}}} & 20 & 2.29 & 1.25 & \num{2.71E+13}  \\
        & 40 & 1.68 & 0.85 & \num{6.73E+13}  \\ 
        & 60 & 1.45 & 0.70 & \num{9.45E+13}  \\ 
        & 80 & 1.33 & 0.60 & \num{1.13E+14} \\
        & 100& 1.26 & 0.53 & \num{1.27E+14}  \\ \midrule
        \parbox[t]{2mm}{\multirow{5}{*}{\rotatebox[origin=c]{0}{$\beta =5$}}}  & 20 & 2.02 & 1.33 & \num{2.63E+13}  \\
        & 40 & 1.38 & 0.92 & \num{6.77E+13} \\ 
        & 60 & 1.16 & 0.74 & \num{9.53E+13}  \\ 
        & 80 & 1.04 & 0.64 & \num{1.13E+14}  \\
        & 100& 0.94 & 0.56 & \num{1.32E+14}  \\ \bottomrule
    \end{tabular}
    \caption{Summary of our lattice results for the gravitational backreaction scenario, including the reheating and backreaction e-foldings as well as the reheating temperature. }
    \label{tab:grav-backreaction-lattice-results}
\end{table}

\subsection{Ricci Reheating via self-interaction backreaction}
\label{subsec:RRviaSelfInteraction}
\begin{figure}
 \includegraphics[width=0.495\textwidth,trim={0.35cm 0.25cm 0.35cm 0.35cm}, clip]{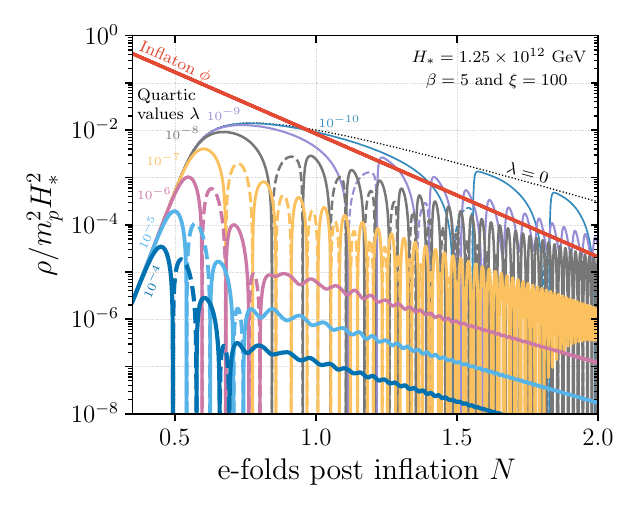}
 \includegraphics[width=0.495\textwidth,trim={0.35cm 0.25cm 0.35cm 0.35cm}, clip]{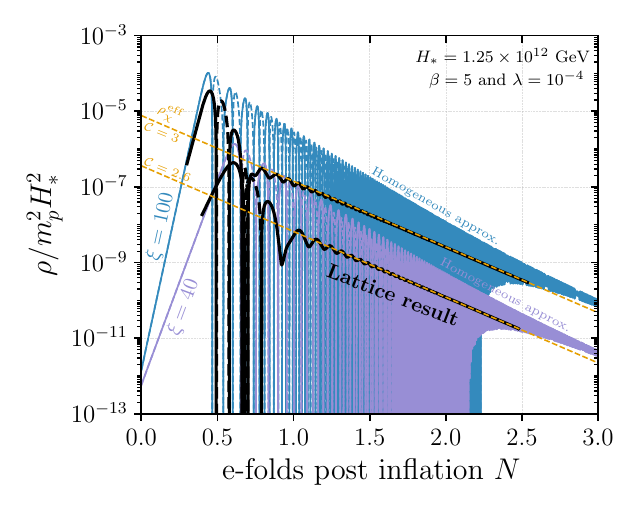}
	\caption{(Left) Energy densities from lattice simulations for several different values of the quartic coupling $\lambda$ for fixed NMC coupling $\xi = 100$. Negative values of the energy density are illustrated using dashed lines on the logarithmic axes. (Right)  Comparison of the lattice simulations ({\bf black}) in this work to the zero-mode approximation ({\color{defblue}\bf blue}, {\color{defpurple}\bf purple}) of Ref.~\cite{Opferkuch:2019zbd}. Also shown is the effective energy density ({\color{cborange}\bf orange dashed}), also from Ref.~\cite{Opferkuch:2019zbd}, rescaled with a correction factor from \cref{tab:latt_corr_factors} to match the lattice results. }
	\label{fig:quartic-comparisons}
\end{figure}

We now turn to the case where $\chi$ has a non-zero quartic self interaction, $\lambda >0$, which will eventually regularize the tachyonic growth of $\chi$. This can be seen in the left-hand panel of \cref{fig:quartic-comparisons}, where we show the energy density in the inflaton and NMC fields for several values of $\lambda$.

\subsubsection{Comparison to previous results} 

Reheating the universe via tachyonic growth of a non-minimally coupled field with a quartic interaction was first proposed in Ref.~\cite{Figueroa:2016dsc}, where the scalar field was identified as the SM Higgs. There it was assumed the tachyonic growth of the field was efficient enough that the background expansion could be neglected. It was later shown in~\cite{Opferkuch:2019zbd} that it takes an $O(1)$ number of e-folds for the backreaction due to the quartic to occur, leading to lower reheating temperatures. However, the analysis in Ref.~\cite{Opferkuch:2019zbd} treated the field homogeneously, so fragmentation effects were neglected. The following analytic approximation to the asymptotic energy density in a massless, non-minimally coupled field was found
\begin{equation}\label{eq:eff-rho}
\rho_\chi^{\rm eff} = \Lambda \, e^{-4(N-N_{\rm min})} \,,
\end{equation}
where $N_{\rm min}$ is the number of e-folds after the end of inflation when $\chi$ reaches the minimum of the effective potential for the first time. Since the field is turned by the quartic coupling soon thereafter, this was used as an approximation for the moment when the tachyonic growth of the field stops. For $w=1$, this time was found to be
\begin{equation}\label{eq:Nmin-Hmin}
N_{\rm min} = \frac{1}{3} \log\left( \frac{H_*}{H_{\rm min}}\right) \,, \hspace{20mm} \left( \frac{H_{\rm min}}{H_*} \right)^{1+\sqrt{\frac{2\xi}{3}}} = \frac{(1 + \sqrt{2})}{24\pi} \left(\frac{\lambda^2}{3 \xi^3}\right)^{1/4}\,.
\end{equation}
The normalization of the energy density $\Lambda$ is given by
\begin{equation}\label{eq:Lambda-result}
\Lambda =  \frac{27\xi^2}{\lambda\,\mathcal{C}(\xi,\lambda) } H_{*}^4 \left( \frac{H_{\rm min}}{H_*} \right)^4\,,
\end{equation}
where in Ref.~\cite{Opferkuch:2019zbd} the {\it fudge} factor $\mathcal{C}$ was fixed identically to 1. As shown in Fig.~\ref{fig:NMC_energy_densities} the gradient energy density contribution is subdominant and hence we expect $\mathcal{C}$ to be an $O(1)$ approximately constant parameter that corrects for the assumption of a homogeneous field assumed in the derivation of \cref{eq:Lambda-result}. The factor $(H_{\rm min}/H_*)^4$ takes into account the dilution in the energy density due to the expansion of the universe before the tachyonic growth is stopped by the quartic. In this work, we fix the value of $\mathcal{C}$ using our lattice simulations, confirming that it is $\mathcal{O}(1)$, see  \cref{tab:latt_corr_factors}. 

\begin{figure}
	\includegraphics[width=0.83\textwidth,trim={0.35cm 0.25cm 0.35cm 0.45cm}, clip]{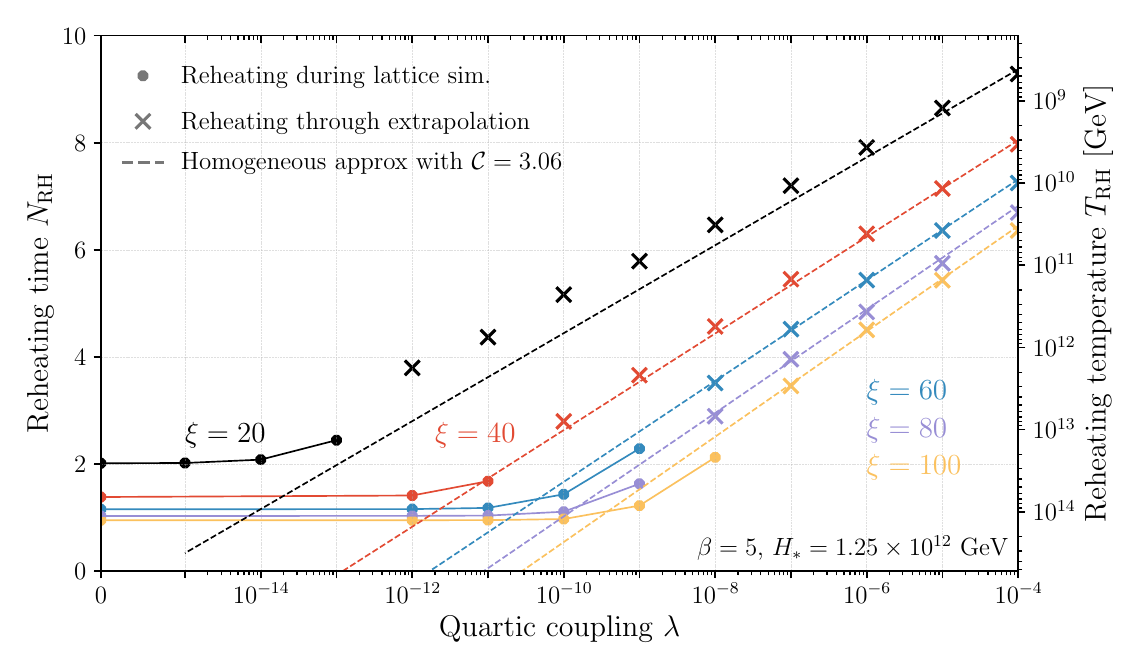}
	\caption{Reheating time in e-folds as a function of the non-minimal and quartic couplings $\xi$ and $\lambda$. Markers indicate points where a lattice simulation was run, where a dot indicates that reheating occurred within the duration of the simulation, while a cross indicates that the reheating temperature is obtained through extrapolation. This extrapolation requires establishing radiation scaling of the averaged NMC fields' energy density which can then be extrapolated to later times to determine the reheating time. Lastly the dashed lines are the predictions using the effective energy density approximation of \cref{eq:eff-rho} (from Ref.~\cite{Opferkuch:2019zbd}) and applying a correction factor $\mathcal{C} = 3.06$, to match with lattice results in the large quartic limit.}
	\label{fig:reheating-time}
\end{figure}

\subsubsection{Reheating time and temperature} 
Using the above expression we also arrive at a closed form result for the number of e-folds required for reheating. 
\begin{align}\label{eq:Nrh-quartic}
    N_\text{RH} &=\log\left[\frac{m_p}{H_*} \frac{\sqrt{\lambda\,\mathcal{C}(\xi,\lambda) }}{3\xi} \left(\frac{H_*}{H_\mathrm{min}} \right)^{4/3}\right]\,.
\end{align}
With this expression and the fact that $H_{\rm RH} \approx H_* e^{-3 N_{\rm RH}}$, we can write the reheating temperature as
\begin{align}
    T_\text{RH} &=\left(\frac{45}{\pi^2 g_{\rm RH}} \right)^{1/4}\sqrt{m_p H_*}~e^{-3 N_{\rm RH}/2} \,\nonumber\\
    &\simeq \SI{5.1E+10}{\GeV} \left(\frac{H_*}{\SI{1.25E+12}{\GeV}}\right)^2\left(\frac{\lambda}{\num{E-4}}\right)^{-\frac{3}{4}+\sqrt{\frac{3}{2\xi}}} \left(\frac{\xi}{100}\right)^{\frac{3}{2} \left(1+ \sqrt{\frac{3}{2\xi}}\right)}\,,
    \label{eq:RHtempRR2}
\end{align}
where we set $\mathcal{C}=3$ and $g_{\rm RH}$ is the number of relativistic degrees of freedom at $T_\text{RH}$, which we take to be $g_{\rm RH} = g_\text{SM} + 1 = 107.75$  \cite{Husdal:2016haj}. Note that for $H_* = \SI{e+11}{\GeV}$, $\lambda = \{\num{e-6}, \num{e-1}\}$ and $\xi = 100$ the fitting functions\footnote{These values are based on private communications with the authors, rectifying typos in the fitting functions presented in \cite{Laverda:2023uqv}.} of \cite{Laverda:2023uqv} yield a reheating temperature of $T_\text{RH} = \{\num{1.2E+10},\SI{9.0e+6}\}\,$GeV \cite{emails, Laverda:2023uqv}, while using our lattice corrected expression we obtain a lower reheating temperature which is more discrepant for larger $\lambda$,  $T_\text{RH} = \{\num{6.2E+9},\num{3.9e+6}\}\,$GeV. 
This discrepancy likely arises from the finite duration of the qdS-KD transition, which becomes more noticeable for larger $\lambda$ and smaller $\xi$. In \cref{fig:reheating-time} our lattice simulations results are shown with a dot or cross, depending on whether reheating occurs during the simulation time or if an extrapolation to later times is required, once radiation scaling of $\chi$'s energy density is established. Also shown (as dashed lines) is the results from the approximate analytic expression in \cref{eq:Nrh-quartic,eq:RHtempRR2}. In particular for sufficiently small $\lambda$, reheating occurs extremely efficiently within two e-folds from the end of inflation. For larger quartic couplings, $\lambda \gtrsim \num{e-8}$, we enter the regime where the parametric scaling of $N_\text{RH}$ as a function of $\xi$ and $\lambda$ is well described by \cref{eq:Nrh-quartic}. Even though the result in \cref{eq:Nrh-quartic} is obtained assuming that the NMC field is homogeneous and that the qdS-KD transition is instantaneous, this equation nevertheless  provides a good description of the lattice results. Namely, the over-estimation of the initial field amplitude is offset by neglecting field fragmentation once the quartic coupling becomes relevant, as well as a slight reduction in the efficiency of the exponential energy growth. The right-hand panel of \cref{fig:quartic-comparisons} illustrates this point with a comparison of our lattice results (thick black lines) with Ref.~\cite{Opferkuch:2019zbd} (thin blue/purple lines). In addition we show the effective energy density $\rho_\chi^\text{eff}$ of \cref{eq:eff-rho} (dashed yellow lines) where $\mathcal{C}$ has been fit to the lattice results using \cref{tab:latt_corr_factors}. It is important to emphasize that even using a mean correction factor\footnote{This is obtained by averaging over all entries appearing in \cref{tab:latt_corr_factors}.} $\mathcal{C} = 3.06$, it is clear that the analytic approximations (dashed lines) provide a good description of the lattice results across a range of parameters. Finally in this figure, departure from the scaling behavior of \cref{eq:Nrh-quartic} (or equivalently the dashed lines), indicates when the quartic is too small for a given size of $\xi$.  This signifies that both the self-interactions and gravitational backreaction 
are important, see for example the $\lambda = \num{e-10}$ or \num{e-9} curves in the left-hand panel of \cref{fig:quartic-comparisons}.

\begin{table}[]
    \centering
    \begin{tabular}{C{1.5cm} |  C{2cm} C{2cm} C{2cm} | C{2cm}} \toprule
         \multirow{2}{*}{$\xi$} & \multicolumn{3}{c |}{Correction factor  
         $\mathcal{C}(\xi,\lambda)$} & \multirow{2}{*}{Avg. $\mathcal{\overline{C}(\xi)}$}  \\ 
         & $\lambda = 10^{-4}$ & $\lambda = 10^{-5}$ & $\lambda = 10^{-6}$ & \\ \midrule
         20 & 2.73 & 3.43 & 4.20 & 3.45 \\
         40 & 2.59 & 2.90 & 3.23 & 2.91 \\ 
         60 & 2.70 & 2.91 & 3.11 & 2.90 \\ 
         80 & 2.80 & 2.91 & 3.07 & 2.92 \\
         100& 2.99 & 3.13 & 3.21 & 3.11 \\ \bottomrule
    \end{tabular}
    \caption{Correction factors as a function of the NMC $\xi$ and the quartic coupling $\lambda$.}
    \label{tab:latt_corr_factors}
\end{table}

\section{Summary and conclusions}
\label{sec:summary}

In this paper, we examine the dynamics of a non-minimally coupled (NMC) scalar spectator field in non-oscillatory inflationary scenarios transitioning from inflation to kination domination (KD) in a finite time, while allowing the field dynamics to be fully inhomogeneous and non-linear. Unlike previous work, which often neglected background expansion during the transition or 
enforced a rigid equation of state (EOS) of $w=1$ during KD, our approach incorporates a precise evolution of this transition from the inflationary phase with $w \gtrsim -1$ to KD with $w \lesssim +1$. We employ a CMB-compatible inflaton potential to simulate a realistic, finite-duration transition to KD, enabling us to investigate the tachyonic instability of the NMC field during this phase. Utilizing lattice simulations, we assess the non-linear dynamics of the NMC field, focusing on how its tachyonic growth is regulated. Our findings provide insights into the conditions necessary for efficient reheating, characterizing the pivotal timescales and energy scales involved. In this work we considered two main mechanism for regulating the tachyonic growth of the NMC field:
\begin{description}[leftmargin=0pt, labelindent=0pt]
    \item[\textbf{Gravitational backreaction}] Scenarios where the NMC field is either free or contains extremely weak interactions reheat extremely efficiently, typically within two e-folds after the end of inflation. Here, the growth in the NMC field is exponential until its energy density becomes a sub-percentage fraction of the inflaton's. At that point the NMC field contributions drive the Ricci scalar to zero, hence regularizing the tachyonic instability. We provide approximate analytic expressions describing this scenario in \cref{eq:BR_expressions}, as well as supporting lattice simulations in \cref{fig:RHtime} where we are able to self-consistently simulate the full system of equations in \cref{eq:inflatonEOM,eq:NMCeom,eq:piadot}. In contrast to the self-interaction backreaction case, here the analytic expressions do not accurately capture the reheating time necessitating the use lattice 
    methods. 
    
    \item[\textbf{Self-interaction backreaction}] Growth of the NMC field is regulated through its own interactions. For the case of a non-zero quartic we validate, using  lattice simulations, the results of Ref.~\cite{Opferkuch:2019zbd} up to a {\it fudge} factor of approximately $\mathcal{C} \approx 3$ in the energy density of the NMC field, see \cref{tab:latt_corr_factors} and \cref{fig:reheating-time}. A key practical result of our work is that future analyses of Ricci Reheating scenarios may simply use the approximate analytic solutions assuming a homogeneous field and instantaneous transition in \cref{eq:eff-rho,eq:Nmin-Hmin,eq:Lambda-result,eq:Nrh-quartic,eq:RHtempRR2}, which we have shown here are excellent approximations to the full lattice result taking into account the correction factors $\mathcal{C}$ given in \cref{tab:latt_corr_factors}.
\end{description}
A very interesting future direction for study on the lattice is the case where the NMC field $\chi$ is identified with the SM Higgs field. However, since the Higgs is necessary coupled to other SM fields, a proper treatment requires a dedicated lattice study that takes into account the running of the Higgs potential as well as non-perturbative production of electroweak gauge bosons. We therefore leave a detailed analysis of this scenario for future work.

\section*{Acknowledgments}

We are very thankful to Adrien Florio for collaboration on related projects and at early stages of this project. We also thank Giorgio Laverda and Javier Rubio for helpful correspondence on their work \cite{Laverda:2023uqv}. DGF is supported by a Generalitat Valenciana grant PROMETEO/2021/083, and by Spanish Ministerio de Ciencia e Innovacion grant PID2020-113644GB-I00. The work of BAS is supported by the STFC (grant No. ST/X000753/1). 

\bibliographystyle{JHEP}
\bibliography{refs}

\end{document}